\documentclass[]{article}
\usepackage{hyperref}
\usepackage{booktabs}
\usepackage{times}
\usepackage{graphicx}
\usepackage{dcolumn} %
\usepackage{bm}      %
\usepackage{fontenc}
\usepackage{rotating}
\usepackage[dvipsnames]{xcolor}
\usepackage{epstopdf}

\usepackage{comment}

\usepackage{amsthm}
\usepackage{amssymb}
\usepackage{amsfonts}
\usepackage{amsmath}
\usepackage{mathrsfs}
\usepackage{mathtools}

\usepackage{subfig}
\usepackage{graphicx}
\graphicspath{{./figures/}}          
\DeclareGraphicsExtensions{.pdf,.jpeg,.png,.jpg,.eps} 
\usepackage{wrapfig}

\usepackage{acro}

\DeclareAcronym{LIF}{
	short = LIF ,
	long  = Leaky Integrate and Fire,
}
\DeclareAcronym{STDP}{
	short = STDP ,
	long  = Spiking Time-Dependent Plasticity,
}
\DeclareAcronym{LTD}{
	short = LTD ,
	long  = Long Term Depression,
}
\DeclareAcronym{LTP}{
	short = LTP ,
	long  = Long Term Potentiation,
}

\date{}
\title{Resonances induced by Spiking Time Dependent Plasticity}
\author{Pau Vilimelis Aceituno}

\begin{document}

\maketitle

\begin{abstract}
	Neural populations exposed to a certain stimulus learn to represent it better. However, the process that leads local, self-organized rules to do so is unclear. We address the question of how can a neural periodic input be learned and use the Differential Hebbian Learning framework, coupled with a homeostatic mechanism to derive two self-consistency equations that lead to increased responses to the same stimulus. Although all our simulations are done with simple Leaky-Integrate and Fire neurons and standard Spiking Time Dependent Plasticity learning rules, our results can be easily interpreted in terms of rates and population codes. 
\end{abstract}

In a more general setting, we would think that forcing a network of neurons to a periodic repetitive signal should make the network represent the signal better. Furthermore,  implemented with the well-known rules presented before using simulations and an analytical approach based on two self-consistency equations relying neural activity and its weights. 

\begin{figure}[h]
	\captionsetup{type=figure}
	\centering
	\includegraphics[width=1\linewidth,trim = {0cm 0cm 0cm 0cm}, clip]{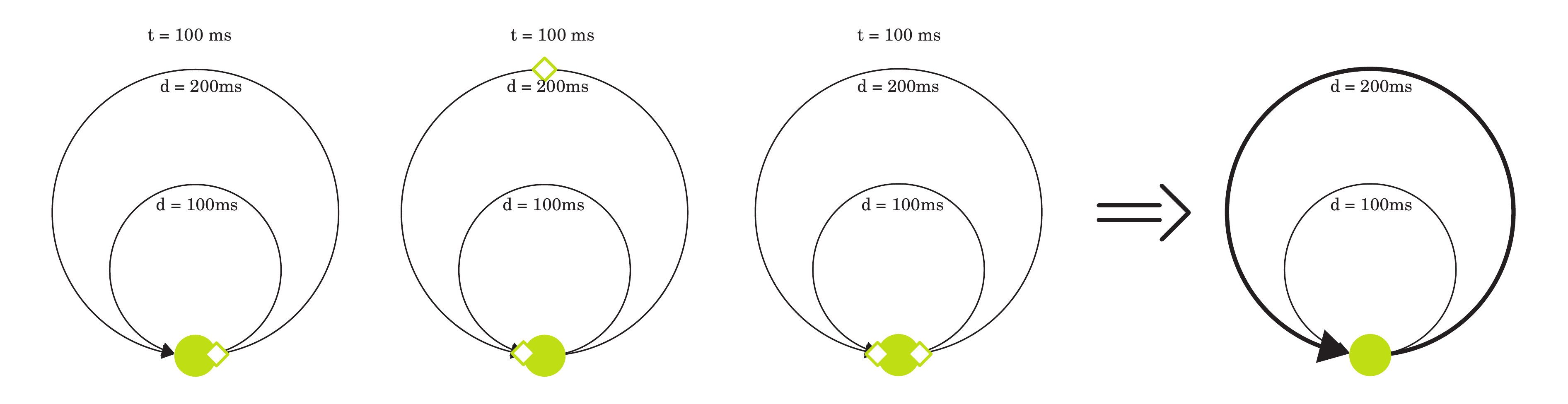}
	\caption[Schema of the emergence of resonances]{\textbf{Schema of the emergence of resonances}: A minimal example of a neural network that induces resonances. The three leftmost schemas show the evolution movement of a spike through the autapses and the rightmost plot the resulting structure of the network. A neuron\label{fig:EmergenceOfResonancesOneNeuron}}
\end{figure}

As a introductory example, we can consider a single neuron with two autapses -- synapses to the neuron itself -- with different delays presented in Fig.~\ref{fig:EmergenceOfResonancesOneNeuron}. One of the autapses has a delay of $100 ms$ while the other has a delay of $200 ms$, on the orders of magnitude od non-myelinated central axons \cite{swadlow2012axonal}. If the neuron is externally forced to fire every $200 ms$, the autapse with delay two has its presynaptic spikes coincide with postsynaptic ones, and thus will get reinforced, while the one with delay of one second would not be affected, on account that the pre- and postsynaptic spikes are too far apart to undergo plasticity. This leads to a neuron with a very strong autapse with delay two, which is effectively a resonant dynamical system with frequency $\frac{1}{2} \ Hz$.

The previous schema is simple to understand but biologically implausible; synapses are typically very fast and autapses are extremely rare, so we will instead use large networks of neurons where the spikes will travel through a large network with short synapses instead of few large synapses. This will require having a network where neurons are active at different points in time, and thus the receptive fields of the neurons must correspond to different phases of the input.

\section{Models}

To be more precise, the model that we will use for simulations and some of our analysis will be the \ac{LIF} model with a refractory period \cite{IFModel}. In this model, the state of a neuron at a given time is described by its membrane potential $v(t)$, which evolves according to the equation
\begin{equation}\label{eq:LIF}
\tau_m \dfrac{d v(t)}{dt} = -(v(t) - v_0) + u(t),
\end{equation}
where $\tau_m = 10 ms$, $v_0 = -70 mV$. $u(t)$ is the input to the neuron at time $t$. When the membrane potential reaches a certain threshold $v_{th} = -50 mV$, the neuron "fires" or "spikes", meaning that it emits a pulse of current -- the spike -- that will be sent to other neurons in the form of a delta function. After firing, the membrane potential is reset to its resting state $v_0$ and kept frozen at this value for a fixed period of time called the refractory period $t_{ref} = 1 ms$. 

The firing of a neuron generates pulses of current that arrive at other neurons, which in turn update their membrane potentials. If neuron $a$ receives the spikes of neuron $b$ we will say that there is a synapse going from the second to the first. The receiving neuron is called postsynaptic and the sending neuron is the presynaptic one. This synapse is characterized by a weight $w_{ab}$ and a delay $d_{ab}$ which correspond, respectively, to the gain and the latency that the pulse of neuron $a$ goes through before arriving at $b$. 

At this point it is useful to discuss the input $u(t)$. As mentioned before, the input to a neuron is often given by spikes from other neurons, which would give us 
\begin{equation}
u(t) = \sum_k w_k\delta(t-t_k),
\end{equation}
where $w_k$ is the weight of the $k$th spike and $t_k$ its arrival time to the neuron. Note that we can also aggregate small contributions from many other neurons forming a smooth function of time or a stochastic variable. In this case, $u(t)$ might be better described by the statistics of this variable, such as the expected value and the variance.

Finally, we shall note that the activity of neuron populations does not need to be described by a set of spike times. In many cases we are interested in a coarser description and then it is better to use variables such as the instantaneous firing rate -- the average number of spikes per time unit -- instead the exact spike times \cite{gerstner2014neuronal}. Furthermore, if we are considering time scales much larger than $\tau_m$, the exact evolution of the inner neuron state is irrelevant and we can focus on the coarse neuron activity which is given by  
\begin{equation}
x(t) = f\left(u(t) \right),
\end{equation} 
where $f$ is a smooth function with lower bounded by zero -- meaning that the neuron cannot fire less than zero spikes -- and upper bounded by $\frac{1}{t_{ref}}$, because the neuron needs $1 ms$ between spikes.

\begin{figure}[h]
	\captionsetup{type=figure}
	\centering
	\includegraphics[width=0.7\linewidth,height=0.25\linewidth,trim = {0cm 0cm 15cm 0cm}, clip]{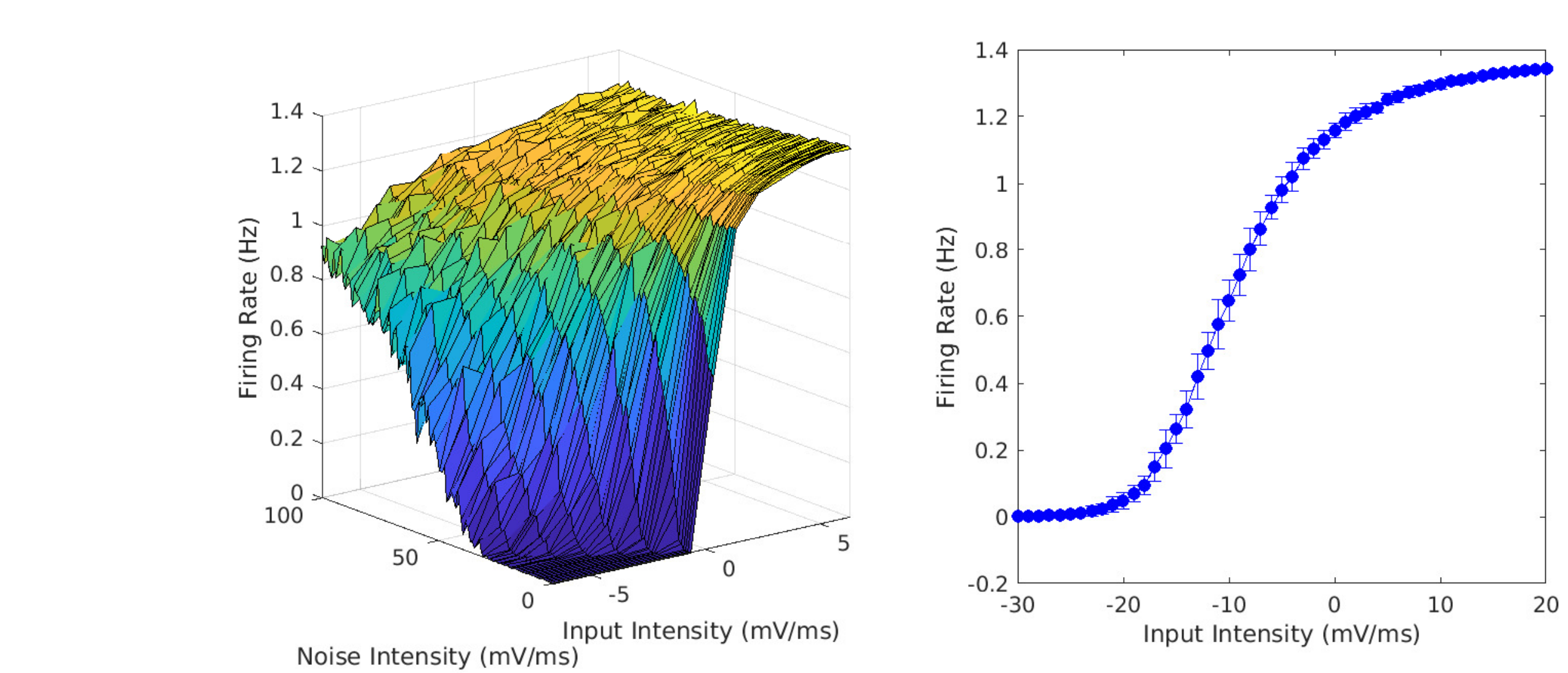}
	\caption[Leaky Integrate-and-Fire as a smooth function]{\textbf{Leaky Integrate-and-Fire as a smooth function}: Here we plot the firing rate of a \ac{LIF} neuron as function of the parameters of a Gaussian input. The left plot presents the average of firing rate for an input presented for 5 seconds and with the two variables being the mean and standard deviation of a Gaussian input. The middle and right plots correspond to cuts through that surface on the respective directions, with the center one being the mean firing rate for a standard deviation of $40 V/s$ and the right one corresponding to varying levels of noise for a mean input of $-5 V/s$.\label{fig:LIFtoRate}}
\end{figure}

\section{Results}

To ease our computations we will use periodic signals with symmetries among neurons and only excitatory synapses. Specifically, we will have the external input being
\begin{equation}
u_n(t) = u_m(t+\Delta t_{mn}) = u_n(t + T)
\end{equation}
where $T$ is the period of the signal, and $\Delta t_{mn}$ simply encodes the difference of timing between neurons $n$ and $m$. Alternatively, we can think of this external input to every neuron as a periodic signal with neurons having different receptive fields.

This external input modifies the activity of the neurons with the equation
\begin{equation}\label{eq:activity}
x_n(t) = f\left(u_n(t) + \sum_{m=1}^{N}\mathbf{W}_{mn}x_m(t)\right)
\end{equation}
where $f$ is a monotonically increasing function with upper and lower bounds as shown in Fig.~\ref{fig:LIFtoRate}. This is just a coarse representation of the \ac{LIF} neurons that represents the probability of a spike at time $t$ --sometimes called the instantaneous firing rate-- when the time scale of the input is much larger than the time scale of the neuron.

\section{Derivation of self-consistency equations}

Given the activities of two neurons we can then apply the STDP equation, which gives us
\begin{equation}\label{eq:weightEvolution}
\Delta w_{mn} = -\gamma(r_n)w_{mn} + \int_{-\infty}^{\infty} \int_{0}^{T} x_m(t) x_n(t + \Delta t)A_\pm(w_{mn}) e^{\frac{\Delta t}{\tau_S}} dt d\Delta t.
\end{equation}
Note that this equation has some parameters that change as the weights evolve, such as the rate $r_n$ or the activities of the neurons. Here we are not interested in the evolution of those, but rather in their final values, so we will treat this equation together with Eq.~\ref{eq:activity} as a system at equilibrium. This implies that $r_n$ is constant, $x_n(t)$ is fixed and $\Delta w_{mn}$ is zero, thus
\begin{equation}
\begin{aligned}
\mathbf{W}_{mn} &= \dfrac{1}{\gamma(r_n)}\int_{-\infty}^{\infty} \int_{0}^{T} x_m(t + \Delta t) x_n(t)A_\pm(w_{mn}) e^{-\frac{|\Delta t|}{\tau_S}} dt d\Delta t.
\end{aligned}
\end{equation}
Furthermore, since we have the $\gamma(r_n)$ homeostatic term we do not need to have bounds on the weights, so we will drop that dependency on $A_\pm$. 

If we notice that the timescale of the STDP is very small so that $\tau_S\rightarrow0$ we can simplify the integral
\begin{equation}
\begin{aligned}
\int_{-\infty}^{\infty}& x_n(t + \Delta t) A_\pm e^{\frac{\Delta t}{\tau_S}}  d\Delta t = \\
&	\int_{-\infty}^{0}x_n(t + \Delta t)A_- e^{\frac{\Delta t}{\tau_S}}  d\Delta t 
+\int_{0}^{\infty}x_n(t + \Delta t)A_+ e^{-\frac{\Delta t}{\tau_S}}  d\Delta t\\
&\approx \kappa_S x_n(t) + \kappa_A\dot{x}_n(t)
\end{aligned}
\end{equation}
where $\dot{x}_n$ is the derivative of $x_n$ and the constants $\kappa_A$ and $\kappa_S$ are constants that account for the symmetric and asymmetric parts of the STDP kernel,
\begin{equation}
\begin{aligned}
\kappa_A &= \int_{0}^{\infty}A_+(w_{mn}) e^{-\frac{\Delta t}{\tau_S}}  d\Delta t
+\int_{-\infty}^{0}A_-(w_{mn}) e^{\frac{\Delta t}{\tau_S}}  d\Delta t\\
\kappa_S &= \int_{0}^{\infty}A_+(w_{mn}) e^{-\frac{\Delta t}{\tau_S}}  d\Delta t
- \int_{-\infty}^{0}A_-(w_{mn}) e^{\frac{\Delta t}{\tau_S}}  d\Delta t,
\end{aligned}
\end{equation}
and therefore we can compute the final weights by
\begin{equation}\label{eq:equilibriumWeight}
\begin{aligned}
w_{mn} &= \lfloor\dfrac{1}{\gamma(r_n)} \int_{0}^{T} x_m(t)\left[\kappa_S x_n(t) + \kappa_A\dot{x}_n(t) (t)\right]dt \rfloor\\
&= \lfloor\beta_S \langle x_m , x_n\rangle + \beta_A \langle x_m , \dot{x}_n\rangle\rfloor
\end{aligned}
\end{equation}
where $\beta_S,\ \beta_A$ correspond to the values of $\kappa_S,\ \kappa_A$ normalized by $\gamma(r_n)$, and $\langle, \rangle$ corresponds to the correlation over the interval $[0,T]$. The brackets $\lfloor\cdot\rfloor$ denote the fact that the weights cannot be negative; even if we were to use balanced networks with inhibitory synapses, the weights would simply decay to zero. 

At this point it is convenient to modify the notation and associate to every neuron a spatial variable $\theta$ such that we can write the input as
\begin{equation}
u_n(t) = u(t,\theta_n) = u(t-c\theta_n)
\end{equation}
where $u(t - c\theta_n)$ is a periodic function that corresponds to the input with different delays due to the neurons position.

Naturally this leads to a similar formulation in terms of the neuron activity,
\begin{equation}
x_n(t) = x(t,\theta_n) = x(t-c\theta_n).
\end{equation}
Here it is important to note that by using this type of notation where the position is only relevant in terms of time we are imposing a spatial symmetry that must be justified. 

In a network where the neurons have different connections, we cannot claim that the activity of one neuron is equal to the activity of another neuron with a shift. This is simply because their inputs from other neurons might differ. Thus, what we are stating here is that not only do neurons obtain the same input with a phase change, but also that 
the matrix $\mathbf{W}$ is built so that
\begin{equation}
\mathbf{W}_{mn} = \mathbf{W}_{m^\prime n^\prime} \quad \forall m - n = m^\prime- n^\prime \mod N.
\end{equation}

Here we shall note that most biological networks of neurons are sparse, meaning that most pairs of neurons are not connected. In small networks, this would imply that the input to a single neuron is not given by its expectation, as we will assume here. The main justification here is the sheer number of connections that every neuron has, which is typically assumed to be on the order of $10.000$ \cite{hawkins2016neurons}. This implies that we can be fairly sure the input to a neuron follows the law of large numbers.

Furthermore, the number of neurons can also be assumed to be large, implying that even if the phases $\theta_n$ are randomly drawn from a uniform probability, the whole phase space will be uniformly covered, again by the law of large numbers.

Another side effect of the symmetries imposed is that we can now describe the activity of the whole network by a single equation
\begin{equation}
x(t) = f\left(u(t) + \int_{0}^{T} \mathbf{w}(\Delta \theta) x(t + c\Delta \theta) d \Delta \theta \right).
\end{equation}
Notice that we are talking about a periodic input, and the phase of each neuron is only given by the phase of that neuron in temporal units that are arbitrary. Thus, we will set $c = 1$ and now we can exchange the integral by a convolution, so that
\begin{equation}\label{eq:consistencyActivity}
x(t) = f\left(u(t) + \left[x\ast \mathbf{w}\right](t) \right)
\end{equation}
where $\ast$ is the convolution operator.

We can also use our new notation and the normalization of $c$ to rewrite Eq.~\ref{eq:equilibriumWeight}
\begin{equation}\label{eq:consistencyWeight}
\begin{aligned}
\mathbf{w}(\Delta \theta) &= \beta_A \int_{0}^{T} x(t) \dot{x}(t + \Delta \theta)dt + 
\beta_S \int_{0}^{T} x(t) x(t+\Delta \theta)dt\\
&= \beta_A\left[x \ast \dot{x}\right](\Delta \theta) + \beta_S\left[x \ast x\right](\Delta \theta).
\end{aligned}
\end{equation}

Eq.~\ref{eq:consistencyActivity} and Eq.~\ref{eq:consistencyWeight} determine the final weights and activity of a neural network subject to the input $u(t)$ after STDP has modified the synapses.

Now we would like to know if they can implement a network adaptation to induce resonance to a given input signal $u(t)$. 

\section{Solution I: Linearization and Sinusoids}

\begin{figure}[h!]
	\captionsetup{type=figure}
	\centering
	\includegraphics[width=1\linewidth,height=0.35\linewidth,trim = {2.3cm 0cm 1cm 0cm}, clip]{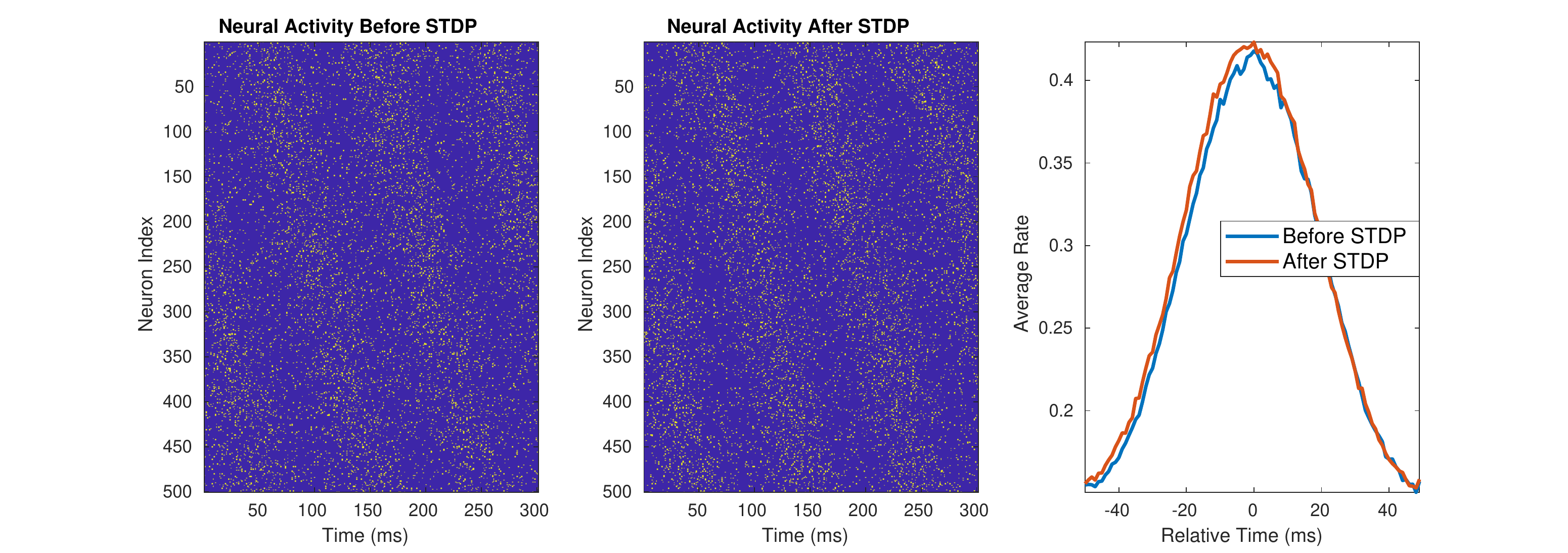}\\
	\includegraphics[width=1\linewidth,height=0.35\linewidth,trim = {2.3cm 0cm 1cm 0cm}, clip]{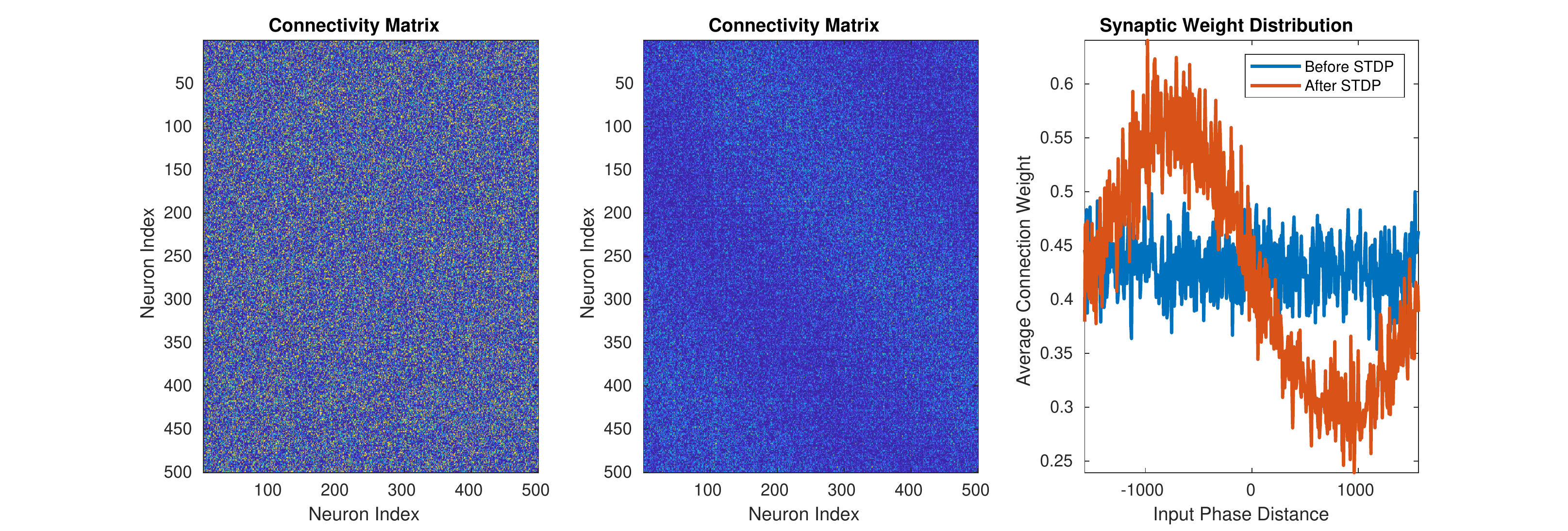}\\
	\caption[Evolution of a circulant peak of activity]{\textbf{Evolution of a circulant peak of activity}: We simulated 400 neurons and feed each one with a sinusoid with phases uniformly distributed across the full circle for all neurons. The upper left and upper center plot correspond to raster plots of neural spikes, and the upper right is the neural activity averaged over ten periods and over all neurons with the phase being centered with respect to their inputs. The lower left and center plots are the matrix weights where brighter color represents higher weights, and the right lower plot is the average vector $\mathbf{w}$. After repeating the input many times, the STDP changes the weights which take the shape of a sinusoid (lower row), and the activity is shifted to be slightly advanced and higher (upper row). The low effect of the weight is simply a result of the parameters of the neurons and weights, which do not allow large weights without generating instabilities through the STDP \label{fig:EvolutionActivitySin}}
\end{figure}

The first problem in Eq.~\ref{eq:consistencyActivity} is that we have the non-linear function $f(\cdot)$, rendering closed-form solutions complicated. The first approach to deal with such situations is to linearize Eq.~\ref{eq:consistencyActivity},
\begin{equation}
x(t) = \varrho \left[u(t) + \int_{0}^{T} \mathbf{w}(\Delta \theta) x(t + c\Delta \theta) d \Delta \theta \right]
\end{equation}
where $\varrho$ is the derivative of $f$ around the mean input to a single neuron,
\begin{equation}
\varrho = \dfrac{\partial f}{\partial u}\left(\bar{u} + \bar{x}\int_{0}^{T} \mathbf{w}(\Delta \theta)d \Delta \theta \right)
\end{equation}
with $\bar{x},\ \bar{u}$ are the mean activity and input respectively. Note that the derivative of $f$ is taken with respect $u$, although it can be taken with respect to any of the inputs to a neuron. The means can be used because the linearization implies that the fluctuations are small, but in any case the main point is that $\varrho$ is constant.

Now that the system consists only of linear equations with convolutions whose solutions are periodic, we can notice that all the operations involved can be written in the Fourier domain,
\begin{equation}
\begin{aligned}
\mathcal{F}\left[x\right](\omega) &= \varrho \left( \mathcal{F}\left[u\right](\omega) + \mathcal{F}\left[x\right](\omega)\mathcal{F}\left[\mathbf{w}\right](\omega)\right)\\
\mathcal{F}\left[\mathbf{w}\right](\omega) &= \left(\beta_A \omega + \beta_S\right)\mathcal{F}\left[x\right]^2(\omega),
\end{aligned}
\end{equation} 
which we can merge to obtain
\begin{equation}
\begin{aligned}
\mathcal{F}\left[x\right](\omega) &= \varrho \left( \mathcal{F}\left[u\right](\omega) + \mathcal{F}\left[x\right]^3(\omega)\left(\beta_A \omega + \beta_S\right)\right) 
\end{aligned}
\end{equation}
This last equation has the advantage that it gives us a direct connection between the input and the activity. More specifically, if we put a single sinusoid with period $T$ as an input, the Fourier transform consists of two deltas at $\omega=\pm2\pi$. Thus, the previous equation is zero everywhere except at the position of the deltas, and there it yields a depressed cubic equation that can be solved analytically.

Even though we have an analytical solution, it is worth noticing that the values of $\varrho,\ \beta_A,\ \beta_S$ must be estimated numerically as they depend nonlinearly on the input statistics to each neuron. Furthermore, there are severe constraints on the parameters of the STDP that arise from the stability of the plasticity and depend on the number of neurons and synapses. Thus, in the simulations presented in Fig.~\ref{fig:EvolutionActivitySin} we observe that there is only a small modification of the input activity. 

That limitation does not imply that our previous results were useless. First, because the results relate the general shape of the STDP kernel to qualitative modifications on the activity, namely the shift in signal phase linked to the strength and sign of $\beta_A$ that arises from the two solutions for $\pm\omega$. Second, because we can obtain the shape of $\mathbf{w}$ as a sinusoid which in our case leans towards the asymmetry and thus gives us weights that are sinusoids with a phase difference in the interval $\left[0, \frac{\pi}{2}\right]$, mostly toward the end of it as shown in Fig.~\ref{fig:EvolutionActivitySin}.

\section{Solution II: Sparsity and Binary Activity}

\begin{figure}[h!]
	\captionsetup{type=figure}
	\centering
	\includegraphics[width=1\linewidth,height=0.35\linewidth,trim = {2.3cm 0cm 1cm 0cm}, clip]{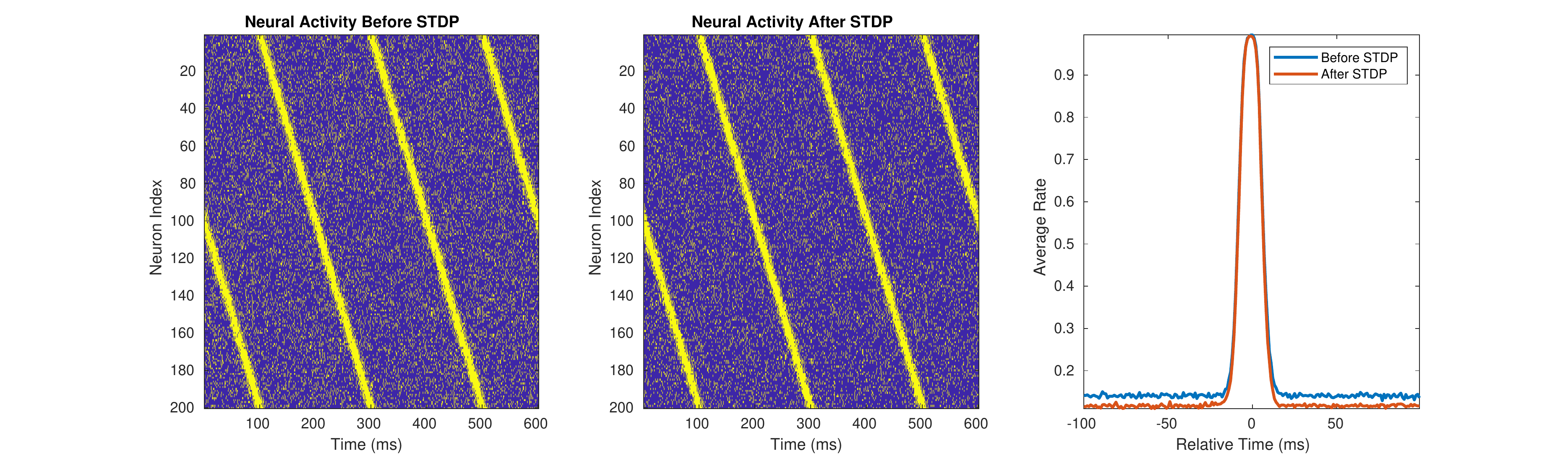}\\
	\includegraphics[width=1\linewidth,height=0.35\linewidth,trim = {2.3cm 0cm 1cm 0cm}, clip]{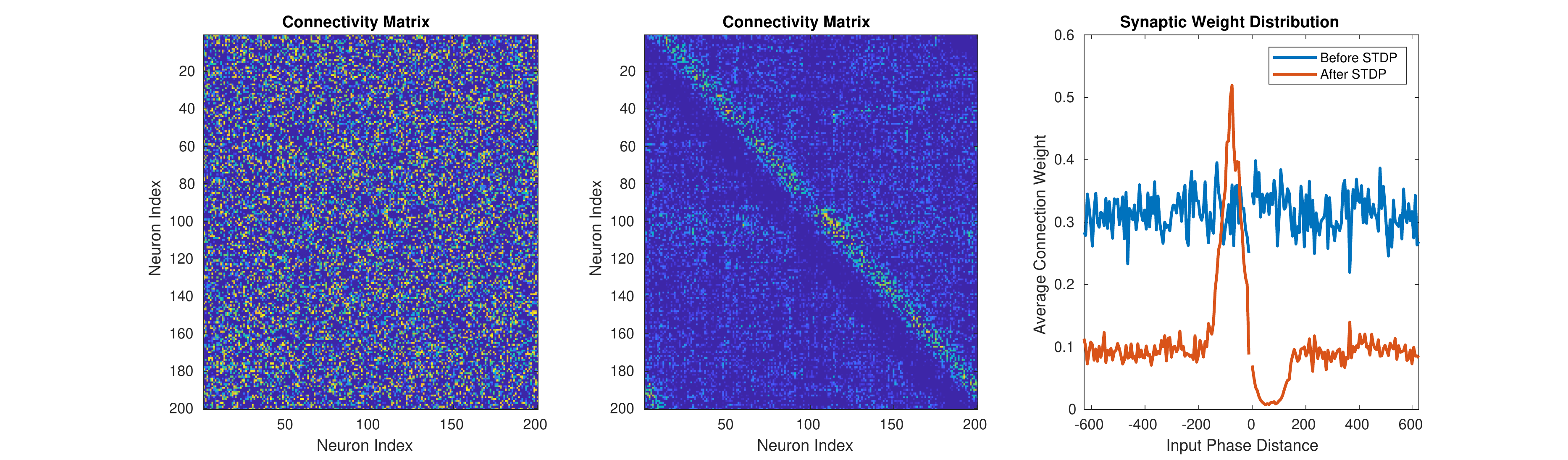}\\
	\caption[Evolution of a circulant peak of activity]{\textbf{Evolution of a circulant peak of activity}: We simulated 500 neurons with random connections and imposed an activity $u(t) \propto \exp\left(-\frac{t^2}{2\tau_u}\right)$ where the time constant $\tau_u$ is on the order of magnitude of the leak time constant so that the activity of neurons with phase $t-c\delta\theta$ affects the activity of neurons at $t$. This activity modifies the initially random network into a circulant-like network, where every neuron connects only to its left neighbors (lower panels), which in turn reduces the noise in the network by decreasing the amount of input at non-active times and increasing the input at the time when the neurons should be active (upper panels).   \label{fig:EvolutionActivityDelta}}
\end{figure}

The second approach to this problem to deal with Eq.~\ref{eq:consistencyActivity} is to take advantage of the particular the nonlinearity of $f(\cdot)$ -- with upper and lower saturation -- and take a signal -- $u(t)$ -- that is encoded as a sequence of very sharp symbols with small overlapping, and then let STDP converge into a neural network where each of those symbols "prepares" the activation of the next. 

The intuition here is that we have a signal consisting in two levels of activity: the peak and the background activity. To improve the signal we only need to push those two levels apart, meaning that the neurons that would be active should have a rate as high possible and those that are not meant to be active should be as silent as possible. This can be seen as a long sequence of symbols where only a small population of neurons should be active for every symbol and with the populations having very little overlap; then, the causal associations promoted by STDP will create a network of neural populations where the populations associated to each symbol have strong connections towards the subsequent population and thus prepare it to receive an input and fire. 

In more practical terms, we are using a simple input which consists of a very narrow peak of current concentrated on very few neurons, and a constant background activity corresponding to the probability of a neuron firing without input. This peak corresponds to a mollified Dirac of activity, which we can plug into Eq.~\ref{eq:consistencyWeight},
\begin{equation}
\begin{aligned}
\mathbf{w}(\theta) = \left\lfloor\beta_A \left[\tilde{\delta}(\theta)\ast a\tilde{\delta}(\theta - \epsilon)  - \tilde{\delta}(\theta)\ast a\tilde{\delta}(\theta + \epsilon)  \right] + \beta_S\left[\tilde{\delta}(\theta)\ast \tilde{\delta}(\theta) + c\right]\right\rfloor
\end{aligned}
\end{equation}
where $\tilde{\delta}$ is the mollified Dirac function, whose derivative is approximated by $a\tilde{\delta}(\theta - \epsilon) - a\tilde{\delta}(\theta + \epsilon)$, with $\epsilon$ being a very small number and $a$ being a scaling that depends on the exact mollification, and the $c$ corresponds to the probability of two neurons randomly firing at similar times thus strengthening their connections by $\beta_S$. Assuming that the mollified version behaves similarly to a Dirac, we can apply the convolutions and obtain
\begin{equation}
\mathbf{w}( \theta) \approx \left\lfloor\beta_Aa\tilde{\delta}(\theta -\epsilon) - 
\beta_A\tilde{\delta}(\theta + \epsilon) + \beta_S\tilde{\delta}(\theta) + c\beta_S\right\rfloor.
\end{equation}

In our case $\beta_A\gg \beta_S >0$, and the weights cannot be negative, the shape of the weights remains constant with value $\beta_Sc$ for most $\theta$, then a peak with height $a\beta_A$ at $-\epsilon$ and a drop to zero at $\epsilon$. Furthermore, with those weights the activity gets reinforced because the neurons at phase $\theta$ promote the activity of neurons at phase $\theta+\delta \theta$, thus the activity increases due to neighboring neurons is mostly arriving at the time when the neuron is getting the high level input. This leads to an increased signal to noise ratio, which is shown in Fig.~\ref{fig:EvolutionActivityDelta}

\section{Discussion}

In this chapter we derived solutions for the evolution of synaptic activity under periodic inputs that use differential Hebbian learning under a self-consistency set-up. In this way we have shown that the features of Echo State Networks that were important in the first part of the thesis can emerge out of synaptic plasticity rules that are well known in the field of neuroscience. In particular, we saw that increased signal-to-noise ratios appear in both the linear and the sparse solution, showing that the signal processing notions can provide neuroscience insight.

The networks proposed here differ from the Echo State Networks used in earlier chapters in the sense that the two solutions presented require very specific receptive fields. Therefore, we cannot use the input as we did in Echo State Networks, but must instead project it into the network in such a way that different neurons are active at different times. 

Although the self-consistency equations used before can in principle be solved numerically, the fact that the parameters $\beta_A,\ \beta_S$ are unknown makes such numerical approaches problematic. However, the limitations outline here are general to theoretical neuroscience, as biological neural networks are typically heterogeneous and estimating the parameters of its components is an arduous task. The results here should therefore be taken qualitatively, in the sense that we would expect STDP to adapt networks of neurons to their input in such a way as to increase the signal to noise-ratio and maybe induce an advancement of the neural activity with respect to the input.

Another remark to this chapter is that we have only proposed the two simplest solutions compatible with Eq.~\ref{eq:consistencyActivity} and Eq.~\ref{eq:consistencyWeight}, but naturally there are other possible options. A straightforward approach would be to increase the multiplicity of the solution, for instance by having multiple isolated sinusoids, or multiple peaks of activity per neuron, in such a way that the general shape of the solution stays the same but neurons have activities composed by multiple receptive fields. 

\section*{Acknowledgement}

P.V.A is supported by the Supported by BMBF and Max Planck Society

\bibliographystyle{unsrt}
\bibliography{STDPInducesResonance.bib}
\end{document}